\documentclass[aps,prl,longbibliography,twocolumn,superscriptaddress]{revtex4-1}
\usepackage{amsmath}
\usepackage{amsthm}
\usepackage{mathrsfs}
\usepackage{graphicx}
\usepackage{color}
\usepackage{tabularx}
\usepackage{array}
\usepackage{multirow}
\usepackage{overpic}
\usepackage[normalem]{ulem}

\DeclareMathOperator{\Tr}{tr}

\newcommand{\rmin}{\mathrm{in}}
\newcommand{\rmout}{\mathrm{out}}

\newcommand{\ket}[1]{|#1\rangle}
\newcommand{\bra}[1]{\langle #1|}

\begin{document}

\title{Tomography and purification of the temporal-mode structure of quantum light}

\author{Vahid~Ansari}
\email{vahid.ansari@uni-paderborn.de}
\affiliation{Integrated Quantum Optics, Paderborn University, Warburger Strasse 100, 33098 Paderborn, Germany}

\author{John~M.~Donohue}
\email{john.matthew.donohue@uni-paderborn.de}
\affiliation{Integrated Quantum Optics, Paderborn University, Warburger Strasse 100, 33098 Paderborn, Germany}

\author{Markus~Allgaier}
\affiliation{Integrated Quantum Optics, Paderborn University, Warburger Strasse 100, 33098 Paderborn, Germany}

\author{Linda~Sansoni}
\affiliation{Integrated Quantum Optics, Paderborn University, Warburger Strasse 100, 33098 Paderborn, Germany}

\author{Benjamin~Brecht}
\affiliation{Integrated Quantum Optics, Paderborn University, Warburger Strasse 100, 33098 Paderborn, Germany}
\affiliation{Clarendon Laboratory, Department of Physics, University of Oxford, Parks Road, OX1 3PU, United Kingdom}

\author{Jonathan~Roslund}
\affiliation{Laboratoire Kastler Brossel, Sorbonne Universit\'{e}s, CNRS, ENS-PSL Research University, Coll\'{e}ge de France; 4 Place Jussieu, 75252 Paris, France}

\author{Nicolas~Treps}
\affiliation{Laboratoire Kastler Brossel, Sorbonne Universit\'{e}s, CNRS, ENS-PSL Research University, Coll\'{e}ge de France; 4 Place Jussieu, 75252 Paris, France}

\author{Georg~Harder}
\affiliation{Integrated Quantum Optics, Paderborn University, Warburger Strasse 100, 33098 Paderborn, Germany}

\author{Christine~Silberhorn}
\affiliation{Integrated Quantum Optics, Paderborn University, Warburger Strasse 100, 33098 Paderborn, Germany}

\begin{abstract}\noindent High-dimensional quantum information processing promises capabilities beyond the current state of the art, but addressing individual information-carrying modes presents a significant experimental challenge. Here we demonstrate effective high-dimensional operations in the time-frequency domain of non-classical light. We generate heralded photons with tailored temporal-mode structures through ultrafast pulse shaping of a parametric downconversion pump. We then implement a quantum pulse gate, enabled by dispersion-engineered sum-frequency generation, to project onto programmable temporal modes, reconstructing the quantum state in seven dimensions. We also manipulate the time-frequency structure by selectively removing temporal modes, explicitly demonstrating the effectiveness of engineered nonlinear processes for mode-selective manipulation of quantum states.\end{abstract}

\maketitle

Photons are critical components of quantum networks and technologies, acting as the natural carrier of quantum information due to their low decoherence, simple transmission, and wide range of encoding possibilities. In particular, high-dimensional encodings offer powerful advantages through an increased information-per-photon capacity~\cite{leach2012secure,zhong2015photon}, complex entanglement structures~\cite{agnew2011tomography,yokoyama2013ultra,roslund2014wavelength,reimer2016generation}, an enhanced resilience to noise and loss~\cite{cerf2002security,vertesi2010closing}, and resource-efficient multi-user networking~\cite{herbauts2013demonstration}. In order to seize these benefits, single photons in clearly distinguishable, accurately controllable, and practically measurable modes are essential to define a high-dimensional quantum alphabet. The spectral and temporal, or time-frequency, photonic degrees of freedom offer an attractive framework for quantum communication and quantum information processing~\cite{tittel98,mower2013high,nunn2013large,lukens2014orthogonal,brecht2015photon,schwarz2016bipartite}.  Unlike polarization and spatial encodings, information encoded in the time-frequency domain is robust through fiber-optic and waveguide transmission, making it a natural candidate for both long-distance quantum communication and compact integrated devices. In particular, broadband temporal modes provide an elegant basis that encodes qudits in intensity-overlapping but field-orthogonal pulses~\cite{brecht2015photon}. Due to their pulsed nature, temporal modes lend themselves to network applications relying on the precise synchronization of multiple parties. Additionally, these temporal modes are a natural choice for physical implementations as they are the eigenbasis of photon pairs emitted from standard parametric downconversion (PDC) sources~\cite{URen2005,roslund2014wavelength}.

To fully exploit the temporal mode structure of quantum light, it is necessary to both control the modal structure of quantum light sources and develop matched mode-selective measurement methods. In order to perform projective measurements onto arbitrary temporal modes, techniques are needed which can identify and remove a specific desired mode from a mixture or superposition. Furthermore, operations on photonic temporal modes must not introduce noise in order to leave the fragile quantum nature of the light intact. Sum-frequency generation with tailored group-velocity relationships and shaped ultrafast pulses provides a capable toolbox for these tasks~\cite{zheng2000spectral,pe2005temporal,eckstein2011quantum,donohue13,reddy2013temporal,allgaier2017bandwidth}. Notably, a sum-frequency process between a weak photonic signal and a shaped strong measurement pulse with matched group velocities has been shown to selectively addresses individual temporal modes~\cite{eckstein2011quantum,reddy2013temporal}.  This process, dubbed the quantum pulse gate (QPG), can be used as a temporal-mode analyzer for communication networks~\cite{zheng2000spectral} or as an add-drop component to build general unitaries and quantum logic gates for a desired temporal-mode basis~\cite{brecht2011quantum,brecht2015photon}. Recent QPG experiments have shown highly efficient and highly selective operations on coherent light pulses~\cite{brecht2014demonstration,kowligy2014quantum,manurkar2016multidimensional,reddy2017engineering,shahverdi2017quantum,reddy2017temporal} and demonstrated its effectiveness as a measurement device for unknown superpositions~\cite{ansari2017temporal} and as a mode-selective photon subtractor~\cite{ra2017tomography}. While some of these works have used weak coherent states~\cite{brecht2014demonstration,manurkar2016multidimensional,ra2017tomography}, no operations on the temporal modes of genuinely quantum light have been demonstrated to date. In continuous-variable quantum optics, homodyne measurements provide inherently mode-selective detection~\cite{polycarpou2012adaptive,qin2015complete,roslund2014wavelength,tiedau2018quantum}, but these techniques do not have the add-drop functionality of the QPG and require knowledge of the underlying photonic quantum state and optical loss to reconstruct the mode distribution.

In this Letter, we show a complete set of tools to generate, measure, and manipulate the temporal-mode structure of single photons with a high degree of control. We orchestrate the modal structure of PDC photon pairs by shaping the pump spectrum. We show that the QPG capably performs projective measurements onto custom temporal modes, with both amplitude and phase sensitivity. We use this functionality to perform a seven-dimensional quantum state tomography of heralded photons and recover their full time-frequency density matrix. We then use the QPG to purify and manipulate the temporal-mode structure of the photons, adjustable through a programmable operation and confirmed through second-order correlation function measurements. We measure high signal-to-noise ratios while operating on quantum light, definitively positioning the QPG as an invaluable resource for pulsed quantum information science.

\begin{figure}[t!]
  \begin{center}
           \includegraphics[width=1\columnwidth]{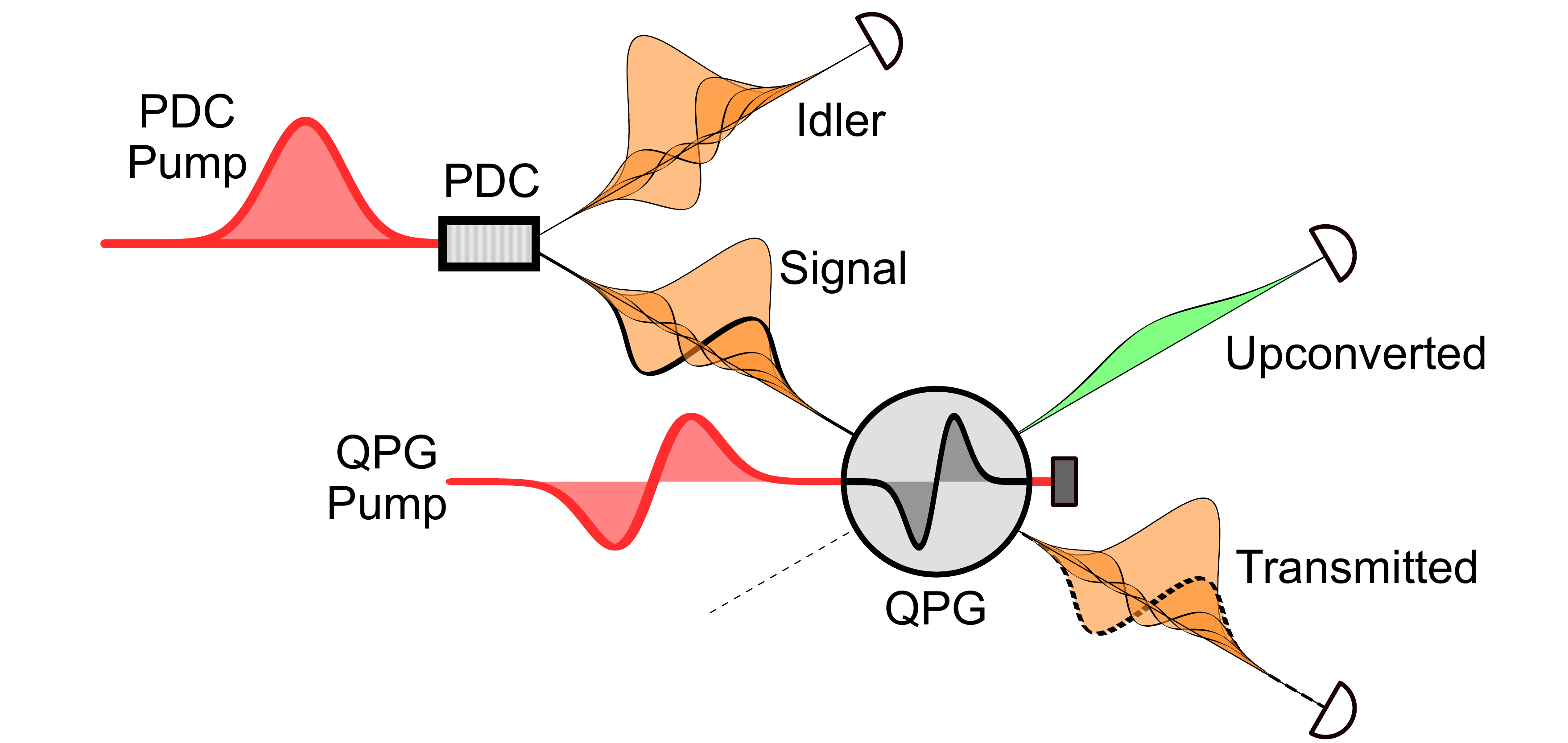}
  \end{center}
  \vspace{-0.25cm}
 \caption{\textbf{Temporal mode selection with a quantum pulse gate.}  The two-photon state resulting from parametric downconversion (PDC) has a multimode structure defined by the pump field and nonlinear phasematching. The quantum pulse gate (QPG) selects a single mode (the first-order Hermite-Gauss for example, in bold) from this superposition and upconverts it to a higher frequency, while the unselected modes transmit unaffected. Changing the shape of the QPG pump changes which temporal mode the QPG selects.}\label{fig:concept}
\end{figure}

We generate photon pairs with a variety of underlying modal structures through parametric downconversion. PDC is a nonlinear process which creates simultaneous signal and idler photons with frequencies $\omega_s$ and $\omega_i$, respectively. The joint spectral amplitude function $f(\omega_s,\omega_i)$ describes the spectral phase and amplitude of the two-photon state, and is determined by the spectral shape of the PDC pump and the dispersive properties of the nonlinear material~\cite{grice1997spectral,kuzucu2005two,URen2005,gerrits2011generation,harder2013optimized}. While the joint spectral amplitude contains a complete description of the state in continuous time-frequency space, an equivalent discrete description can be obtained from the Schmidt decomposition~\cite{law2000continuous}, which re-expresses it in terms of orthonormal modes with normalized Schmidt coefficients $\gamma_k$ as \begin{equation}f(\omega_s,\omega_i)=\sum_k\sqrt{\gamma_k}\,\psi_k(\omega_s)\phi_k(\omega_i).\label{eq:PDCschmidt}\end{equation} For a Gaussian joint spectral amplitude, the eigenmodes are given by Hermite-Gaussian (HG) functions, as sketched in Fig.~\ref{fig:concept}. Notably, these modes have overlapping intensities and therefore cannot be isolated or measured with standard frequency filtering~\cite{shahverdi2017quantum}.

In this discretized picture, the density matrix $\rho_{si}$ containing the complete time-frequency description of the two-photon state can be written simply as \begin{equation}\rho_{si}=\sum_{i,j}\sqrt{\gamma_i\gamma_j}\ket{\psi_i\phi_i}\bra{\psi_j\phi_j},\end{equation} where $\ket{\psi_i}$ and $\ket{\phi_i}$ are the signal and idler photon states, respectively, defined by the corresponding temporal modes. By detecting the idler photon in a time-frequency insensitive manner, the state of the signal photon collapses to $\rho_s={\sum_i\gamma_i\ket{\psi_i}\bra{\psi_i}}$, with a purity of $P=\Tr{\rho_s^2}=\sum_i\gamma_i^2$. In the low-gain regime, the purity is directly related to the second-order autocorrelation function (i.e. the marginal $g^{(2)}$) as $g^{(2)}=1+P$~\cite{christ2011probing}. This provides an experimentally accessible measure of the underlying modal structure, directly probing the decomposition of the joint spectral amplitude independent of the individual mode shapes.

We implement a mode-selective QPG through sum-frequency generation, a nonlinear process which couples input frequencies $\omega_\rmin$ to output frequencies $\omega_\rmout$, according to a mapping function $\xi(\omega_\rmin,\omega_\rmout)$. This function is given by the product of the complex spectrum of the QPG pump $\alpha(\omega_\rmout-\omega_\rmin)$ and the phasematching of the material $\Phi(\omega_\rmin,\omega_\rmout)$. If the input signal and the QPG pump have the same group velocity, the phasematching can be written as a function of only the output frequency, i.e. $\Phi(\omega_\rmin,\omega_\rmout)\approx\tilde{\Phi}(\omega_\rmout)$. For a sufficiently long interaction length, the output frequency range is much narrower than the input~\cite{allgaier2017bandwidth}, and the contribution of the QPG pump spectrum can be approximated as a function of only the input frequency, $\alpha(\omega_{\rmout}-\omega_\rmin)\approx\tilde{\alpha}(\omega_\rmin)$. In this limit, the mapping function $\xi(\omega_\rmin,\omega_\rmout)$ becomes separable, and we can describe the QPG interaction as a single-mode broadband beamsplitter coupling an input temporal mode defined by $\tilde\alpha(\omega_\rmin)$ to the upconverted mode defined by $\tilde\Phi(\omega_\rmout)$, while transmitting all orthogonal temporal modes unaltered ~\cite{eckstein2011quantum,reddy2013temporal}. This is schematically depicted for a QPG set to the first-order Hermite-Gauss mode in Fig.~\ref{fig:concept}. By measuring a photon in the upconverted mode, we implement a projective measurement onto a temporal mode that can be freely chosen through standard pulse shaping of the QPG pump~\cite{ansari2017temporal}.

The group-velocity matching condition can be met in periodically poled lithium niobate (PPLN) waveguides, which also provide the spatial confinement necessary for long nonlinear interaction lengths. In our experimental setup, detailed in the Supplemental Material, we make use of type-II group-velocity matching between a 1540-nm photonic input and an 876-nm QPG pump in a homemade 17-mm PPLN waveguide, as in Refs.~\cite{brecht2014demonstration,ansari2017temporal}. We measure upconverted output pulses at 558~nm with a 61-pm (59~GHz) bandwidth (full-width at half-maximum), significantly narrower than the 4.9-nm (620~GHz) bandwidth of the input photons. Although similar conditions can be met in other materials using near-degenerate processes~\cite{kowligy2014quantum,manurkar2016multidimensional,reddy2017engineering}, our scheme avoids the challenge of isolating the single-photon signal from the second harmonic of the QPG pump.

\begin{figure}[t!]
  \begin{center}
    \includegraphics[width=1\columnwidth]{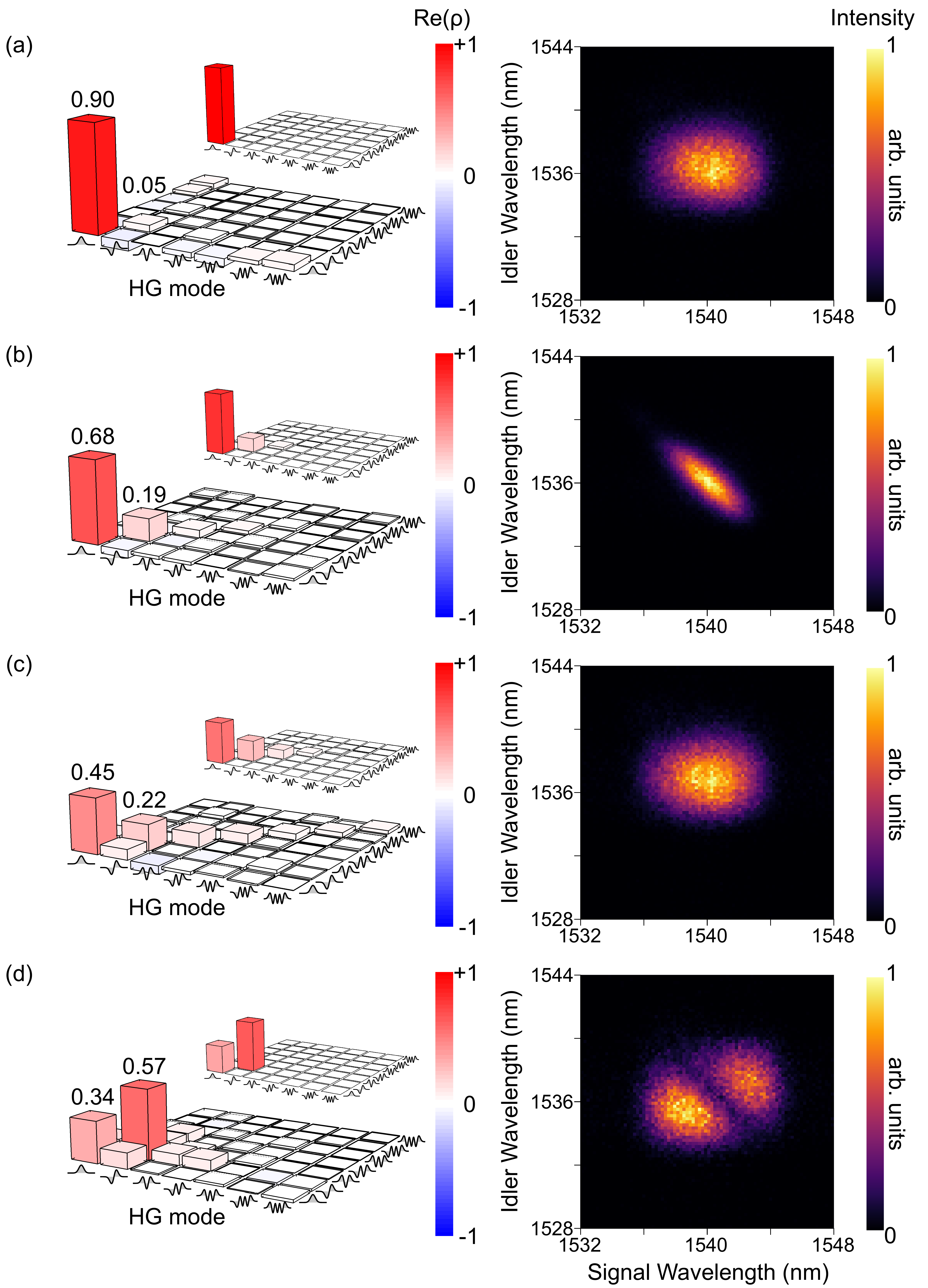}
  \end{center}
  \vspace{-0.25cm}
 \caption{\textbf{Joint spectral intensities and reconstructed temporal-mode density matrices.}The real part of the seven-dimensional one-photon temporal-mode tomographically reconstructed density matrices (left), joint spectral intensities (right), and theoretically expected density matrices (inset) for four PDC states: (a) a separable PDC state, (b) a PDC state with spectral anti-correlations from a narrow-bandwidth pump, (c) a PDC state with spectral phase correlations from a chirped pump, and (d) a PDC state pumped with a higher-order mode. The values of the first two diagonal entries are explicitly labelled above the density matrix. Imaginary components of the reconstructed density matrices are small and found in the Supplemental Material.}\label{fig:jsitomo}
\end{figure}

We use spatial-light-modulator-based pulse shapers to define both the spectral amplitude and phase of the PDC and QPG pump pulses~\cite{weiner2000femtosecond,monmayrant2010newcomer}. With this flexibility in hand, we selected four PDC states to illustrate the versatility of the QPG. The joint spectral intensity $|f(\omega_s,\omega_i)|^2$ for each is shown of the right side of Fig.~\ref{fig:jsitomo}, as measured with dispersive time-of-flight spectrometers~\cite{avenhaus2009fiber,gerrits2011generation}. Firstly, we set the PDC pump bandwidth such that the generated two-photon state is nearly spectrally single-mode~\cite{harder2013optimized}, as seen in the nearly separable joint spectral intensity in Fig.~\ref{fig:jsitomo}a. A singular value decomposition of the joint spectral intensity predicts a purity of 0.995, but measured $g^{(2)}=1+P$ (corrected for detector dark counts) corresponds to a significantly lower purity of ${0.929\pm0.008}$, potentially due to remaining phase correlations or degenerate background processes.

By shaping the QPG pump to project onto a set of Hermite-Gauss spectral shapes, we expect significantly higher upconversion probabilities for the lowest-order Gaussian mode. We find that, when measuring in coincidence with an idler detection, the Gaussian projection indeed provides more counts than the first-order Hermite-Gaussian projection by a factor of 19.3 (12.8~dB), with even stronger suppression for higher-order modes. This demonstrates simultaneously the high mode separability of our device and the single-mode character of our PDC state. With a coherent-state input signal from a commercial pulse shaper instead of PDC photons, the suppression factor increases to 111 (20.5~dB). An in-depth characterization of the mode selectivity of this device with classical light can be found in Ref.~\cite{ansari2017temporal}. The upconverted signal is cleanly separated from all background sources, even for a PDC-generated average photon number of $\langle{n}\rangle\approx0.16$. The signal-to-noise ratio, including noise from detector dark counts, scattered strong laser light, and competing nonlinear noise processes in the poled waveguide~\cite{pelc2011long}, is over 70:1 without heralding and increases to over 900:1 when gated by an idler detection.

While joint spectral intensity measurements provide important information about the two-photon PDC state, they potentially hide significant information about the spectral phase to which mode-selective measurement would be sensitive. To demonstrate the effectiveness of the QPG for quantum state characterization, we reconstruct the density matrix of the signal photons, as seen on the left-hand side of Fig.~\ref{fig:jsitomo}. By shaping the QPG pump, we project onto the first seven Hermite-Gauss temporal modes as well as a tomographically complete set of superpositions, totalling 56 measurements~\cite{wootters1989optimal,bandyopadhyay2002new}. The time-frequency waveforms chosen span eight mutually unbiased seven-dimensional bases, and are sketched in the Supplemental Material. The density matrices were then reconstructed from the heralded counts in the upconverted mode using a maximum-likelihood approach~\cite{altepeter2005photonic}. As the tomography measurements are made on one photon of a PDC pair, we expect to reconstruct mixed density matrices with purities consistent with the measured $g^{(2)}$. For the separable PDC state of Fig.~\ref{fig:jsitomo}a, we reconstruct a density matrix with a purity of $\mathrm{Tr}(\rho^2)=0.896\pm0.006$, lower than the expected value of ${0.929\pm0.008}$. Discrepancies between the tomographically reconstructed purities and the $g^{(2)}$ values arise from slightly diminished mode selectivity for the higher-order projections~\cite{ansari2017temporal}, to which characterization of single-mode behaviour is particularly sensitive.

\begin{figure}[t!]
  \begin{center}
    \includegraphics[width=1\columnwidth]{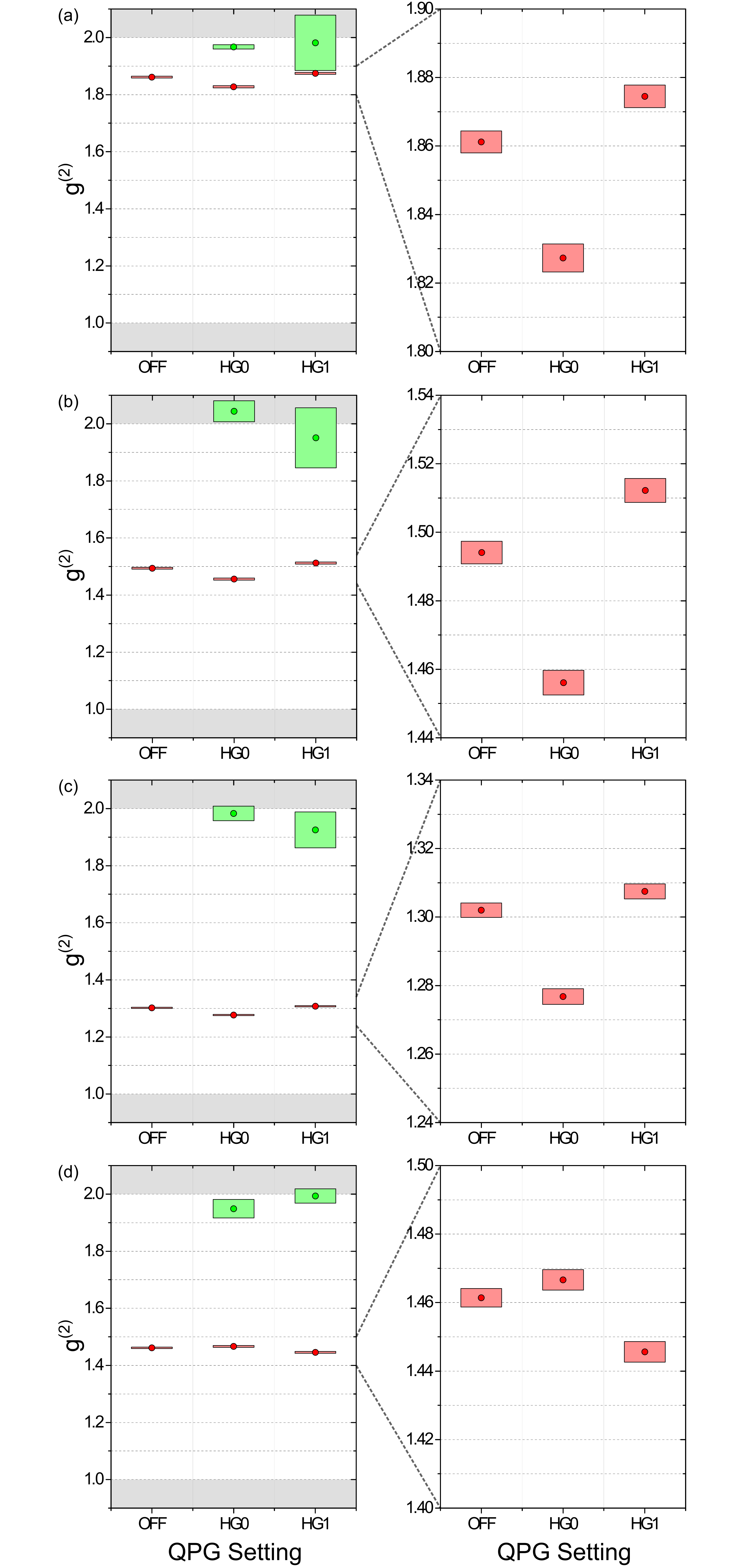}
  \end{center}
  \vspace{-0.25cm}
 \caption{\textbf{Second-order autocorrelation functions of transmitted and upconverted photons.} The marginal $g^{(2)}$s of the upconverted (green) and transmitted (red) PDC photons are shown for the four PDC states corresponding to Fig.~\ref{fig:jsitomo}a-d with the QPG pump pulse delayed relative to the signal photons (`OFF') and shaped to the first two Hermite-Gauss temporal modes (`HG0' and `HG1').  The right side of the figure shows the same data rescaled to highlight the changes in the $g^{(2)}$ of the transmitted photons. The data presented is dark-count background subtracted and the error bars are found assuming Poissonian noise.}\label{fig:g2s}
\end{figure}

Next, we increase the number of modes present in the PDC state in three different ways, and show that the QPG measurements are sensitive to all three. First, we narrow the bandwidth of the PDC pump to produce a multimode PDC state with spectral intensity anticorrelations. The inseparability of this system can be seen directly in the anticorrelations of the joint spectral intensity as well as the additional components of the reconstructed density matrix in Fig.~\ref{fig:jsitomo}b. The purity of the reconstructed density matrix is found to be $0.523\pm0.008$, which matches the $g^{(2)}$-inferred purity of $0.528\pm0.009$.

Intensity correlations are not the only available avenue for increasing the mode number of a PDC state. By adding quadratic spectral phase (chirp) to the PDC pump, we introduce phase correlations between the signal and idler photons. Note that this phase does not affect the joint spectral intensity, as seen in Fig.~\ref{fig:jsitomo}c. However, the added phase drastically decreases the $g^{(2)}$, with a measured purity of $0.327\pm0.005$. Through tomography, we find that the QPG measurements are also sensitive to this phase, with significantly more diagonal elements in the density matrix and a reconstructed purity of $0.317\pm0.005$, similar to the $g^{(2)}$-inferred purity. This result explicitly demonstrates the limitations of spectral intensity measurements for benchmarking pure single photons and the necessity of spectral phase control. More details on PDC with a chirped pump can be found in the Supplemental Material.

In each of the previous cases, the primary temporal mode of the PDC state is approximately Gaussian, with higher-order contributions falling off exponentially and no finite cutoff. As a final example, we demonstrate control over the modal composition within a restricted subspace. We produce a state with contributions from principally two temporal modes by shaping the PDC into the first-order Hermite-Gauss function, as seen in Fig.~\ref{fig:jsitomo}d, which is expected to produce photon pairs in the time-frequency Bell state~\cite{brecht2015photon}. The reconstructed density matrix from the QPG measurements shows that the modal content of the PDC state is mainly confined to these two modes, and the purities inferred from the $g^{(2)}$ and the tomography for this state are, respectively, $0.498\pm0.006$ and $0.531\pm0.004$, consistent with half of a highly entangled qubit pair. The imbalance between the first two modes can be attributed to a non-ideal group-velocity relationship between the signal and idler in the PDC process (i.e. a non-45-degree phasematching angle~\cite{harder2013optimized}), and is consistent with the density matrix expected from the joint spectral intensity.

In addition to being useful as a measurement tool, the QPG in combination with the PDC source can also be used as a source of single-mode photons by isolating one mode from all others~\cite{brecht2015photon}. To demonstrate this state purification, we measure the $g^{(2)}$ of the upconverted photons with the QPG pump in the first two Hermite-Gauss modes, as shown in Fig.~\ref{fig:g2s}. If the QPG isolates a single mode from the input mixture, the upconverted photons themselves will be highly pure. Indeed, the $g^{(2)}$ of the upconverted light confirms a purity of at least 0.9 for both the zeroth- and first-order HG modes, regardless of the PDC state under interrogation. For example, for the correlated spectral intensity of case (b), the $g^{(2)}$ of the upconverted light when the Gaussian mode is selected is $1.95\pm0.04$, which increases to $2.04\pm0.04$ after dark-count subtraction. The purity of the upconverted light remains high when the first-order HG mode is selected. The high $g^{(2)}$ values measured here conclusively show both that the QPG indeed selects a single mode and that the upconverted mode retains the thermal photon statistics of PDC, with very little noise introduced by the process.

Finally, we show through the $g^{(2)}$ that the modal structures of the transmitted photons are significantly altered by the QPG. If a mixture of modes is dominated by one mode, partially removing that mode from the mixture will increase the mixedness of the remaining distribution, akin to the Procrustean method of entanglement concentration~\cite{bennett1996concentrating}. For the decorrelated PDC state of case (a), we measured the conversion efficiency through the depletion of the transmitted signal as approximately 22\%, limited by the nonlinear interaction strength and the available QPG pump power. This partial removal of the primary mode indeed results in a significant decrease in the $g^{(2)}$ of the unconverted transmitted signal photons, as seen on the right-hand side of Fig.~\ref{fig:g2s}, consistent with the efficiency measured from the input depletion. Conversely, removing the first-order HG mode removes amplitude from the secondary Schmidt coefficient, which increases the relative amplitude of the primary Schmidt mode. This is seen in cases (a-c) to increase the overall purity of the transmitted photon state, demonstrating that the QPG can act as a temporal mode cleaner even for the non-converted photons. In case (d), the first-order HG mode is present in a larger proportion than the Gaussian component, and the opposite trend is seen. This result directly demonstrates that the QPG can be used to remove modal components from a single-photon state. QPG efficiencies above 80\% have been demonstrated with classical light~\cite{brecht2014demonstration,manurkar2016multidimensional,reddy2017engineering}, and schemes to reach unit efficiency have been shown with double-pass configurations~\cite{reddy2014efficient,reddy2017temporal}, which combined with this result pave the way for mode-selective add/drop functionality.

We have shown that the quantum pulse gate can be used to directly manipulate and measure the temporal modal structure of single-photon states. By projecting over a complete set of temporal modes and superpositions, we reconstructed seven-dimensional temporal-mode density matrices for PDC photons with a variety of modal structures.  We have demonstrated that the output of the pulse gate is nearly completely purified regardless of the input, positioning the quantum pulse gate as a powerful tool for photonic quantum state engineering.  We have also demonstrated through changes in the second-order autocorrelation function that the quantum pulse gate modifies the modal structure of the input photons, establishing the QPG as a novel device for both entanglement concentration and state purification.  Future work will focus on improving the efficiency and extending the accessible dimensionality of the quantum pulse gate to fully realize its potential for time-frequency mode-selective measurement, as a conversion interface and add/drop device for temporally encoded quantum networks, and as a platform for high-dimensional quantum state characterization.

\begin{acknowledgments}We thank T.~Bartley, D.~V.~Reddy, and J.~Tiedau for fruitful discussions. This research has received funding from the Deutsche Forschungsgemeinschaft (DFG) via Sonderforschungsbereich TRR 142, the Gottfried Wilhelm Leibniz-Preis, and from the European Union’s (EU) Horizon 2020 research and innovation program under Grant Agreement No. 665148. J.M.D. gratefully acknowledges support from Natural Sciences and Engineering Resource Council of Canada (NSERC).\end{acknowledgments}



%

\clearpage
\onecolumngrid
\appendix

\textit{\textbf{Supplemental Material}}

\begin{figure}[b!]
  \begin{center}
           \includegraphics[width=0.5\columnwidth]{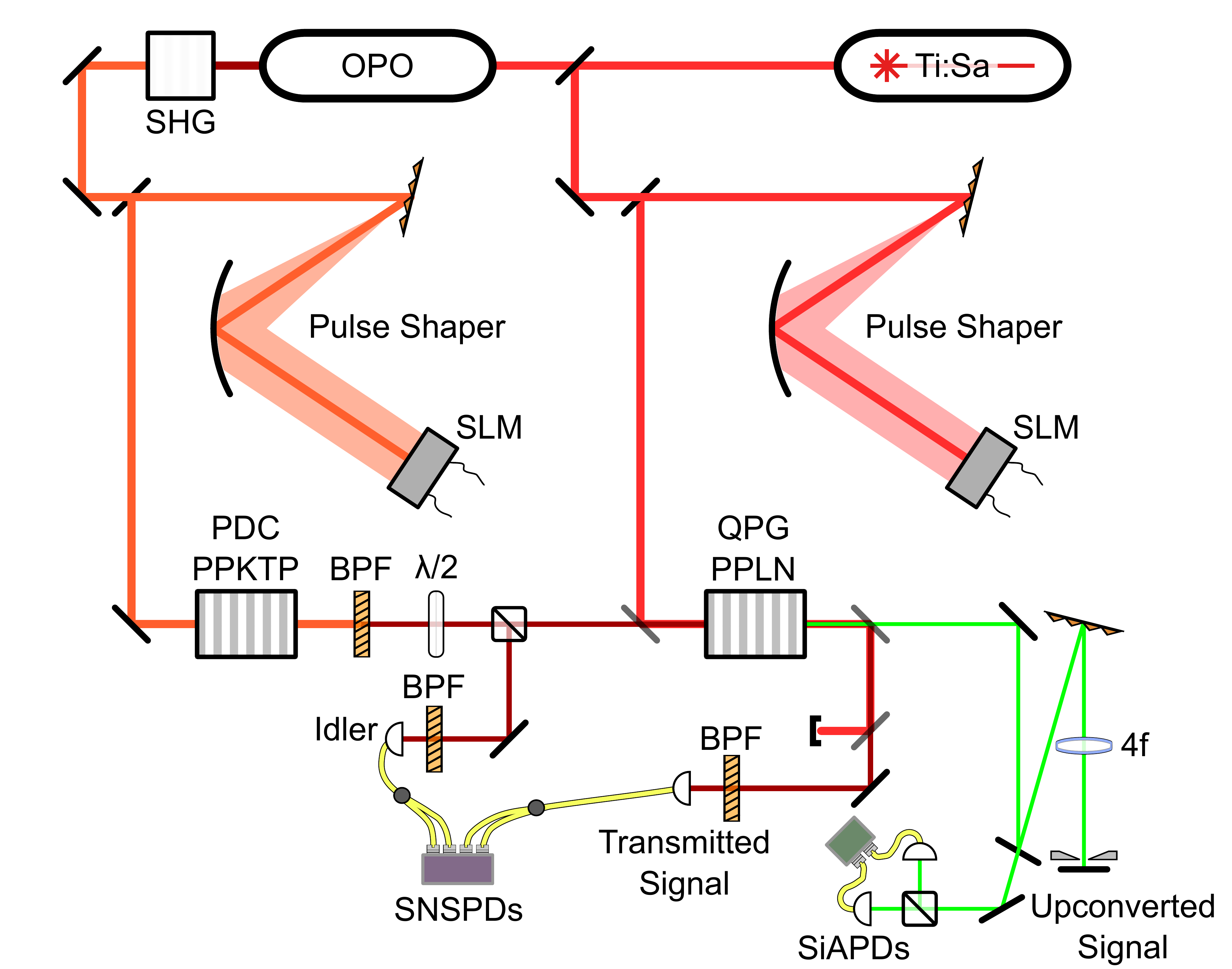}
  \end{center}
  \vspace{-0.25cm}
 \caption{\textbf{Experimental setup.} We create photon pairs through type-II PDC in an 8-mm PPKTP waveguide. By shaping the bandwidth and spectral phase of the PDC pump with a spatial light modulator (SLM) in a 4f line~\cite{weiner2000femtosecond,monmayrant2010newcomer}, we can control the effective mode number of the generated photon pairs.  The PDC pump is removed with a bandpass filter (BPF) and the photon pair is split with a polarizing beamsplitter (PBS). The signal photon is then coupled into a 17-mm PPLN waveguide acting as a quantum pulse gate (QPG), with a QPG pump shaped in both phase and amplitude by another SLM.  A series of dichroic mirrors and a 4f line are used to split the upconverted and transmitted photons from the leftover QPG pump, and all photon paths are coupled into single-mode-fiber beamsplitters to measure second-order correlation functions.}\label{fig:setup}
\end{figure}

\begin{figure}
  \begin{center}
    \includegraphics[width=0.625\columnwidth]{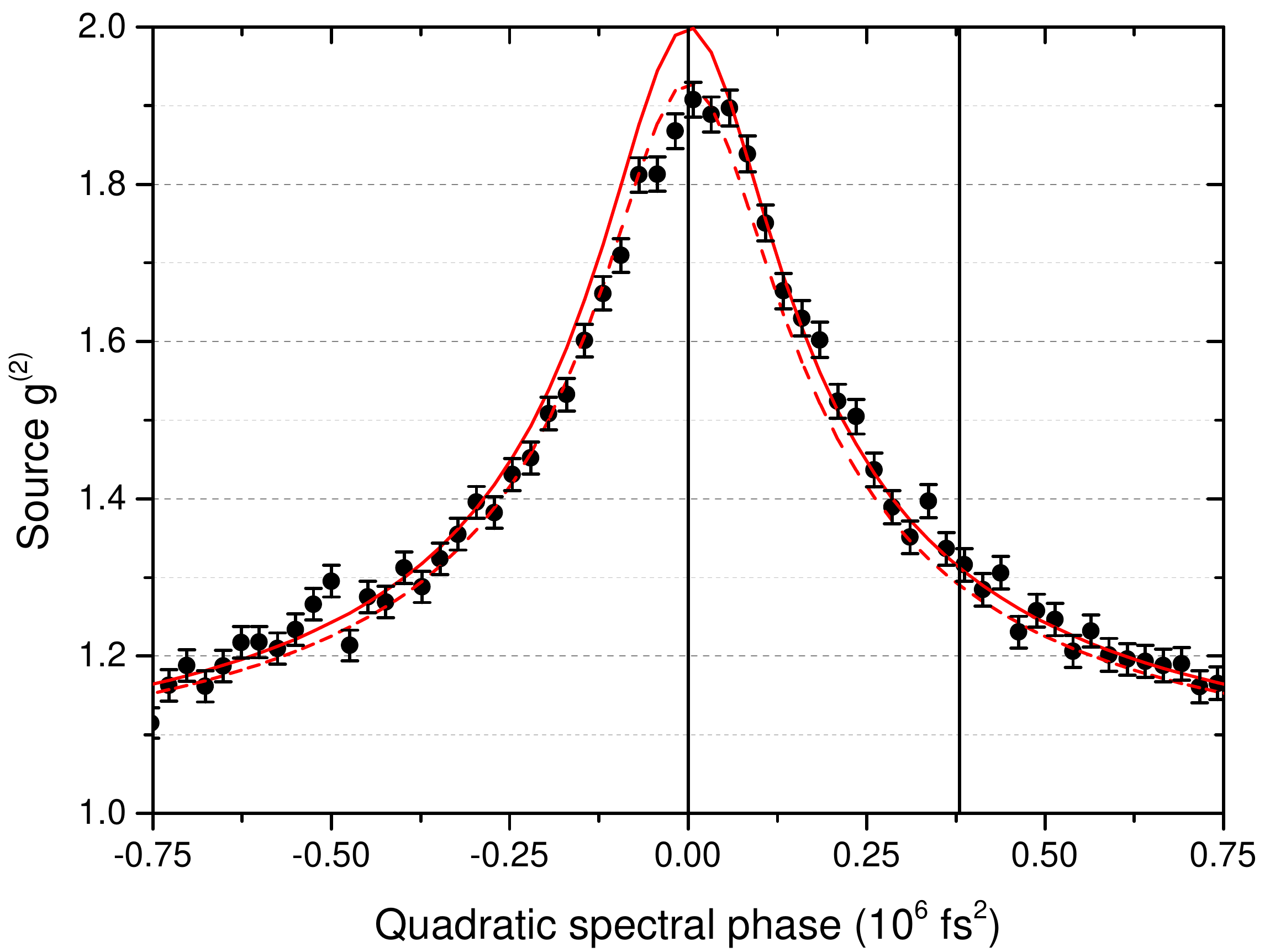}
  \end{center}
  \vspace{-0.25cm}
 \caption{\textbf{PDC source purity as spectral phase added.} The purity of the PDC source, as measured from the marginal $g^{(2)}$ of the signal photon with the QPG off, as a function of quadratic spectral phase of the form $e^{iA(\omega-\omega_0)^2}$ on the PDC pump. Five seconds of data were taken per spectral phase setting. The thick black lines represent the chirp values used for a high-purity PDC state (Case `a') and a highly multimode PDC state (Case `c'). The solid red curve is the theoretical expectation of Eq.~\ref{eq:chirppurity}, and the dashed red curve is the same curve with Poissonian background equivalent to 4\% of the total count rate added to match the peak $g^{(2)}$ of $1.929$ measured in the experiment.}\label{fig:chirpedsource}
\end{figure}

\begin{figure}
  \begin{center}
    \includegraphics[width=1\columnwidth]{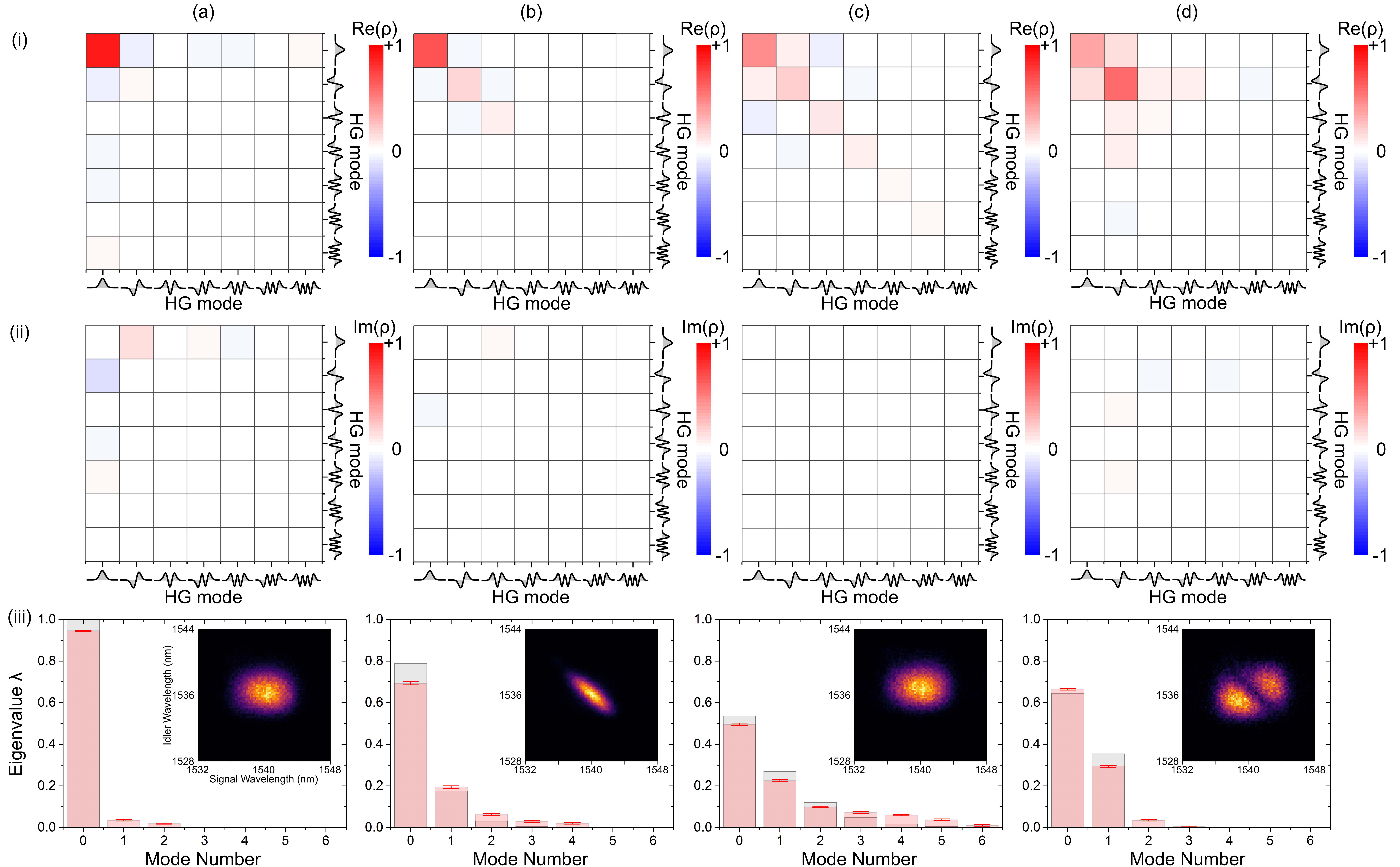}
  \end{center}
  \vspace{-0.25cm}
 \caption{\textbf{Reconstructed temporal-mode density matrices and joint spectral amplitudes.} The real (i) and imaginary (ii) parts of the reconstructed signal-photon density matrices for (a) a spectrally decorrelated PDC state, (b) an intensity-correlated state, (c) a phase-correlated state, and (d) an HG1-pumped state. The eigenvalues ($\sum\lambda=1$) of these density matrices are shown in red in (iii), with the error bars found from Monte Carlo simulations assuming Poissonian noise. The expected one-photon density matrices from the joint spectral intensities (inset) are all diagonal with eigenvalues obtained from the singular value decomposition, as seen in gray assuming a flat phase for cases (a) and (b) and the programmed phase in (c) and (d).}\label{fig:tomoReAndIm}
\end{figure}

\begin{figure}
  \begin{center}
    \includegraphics[width=0.75\columnwidth]{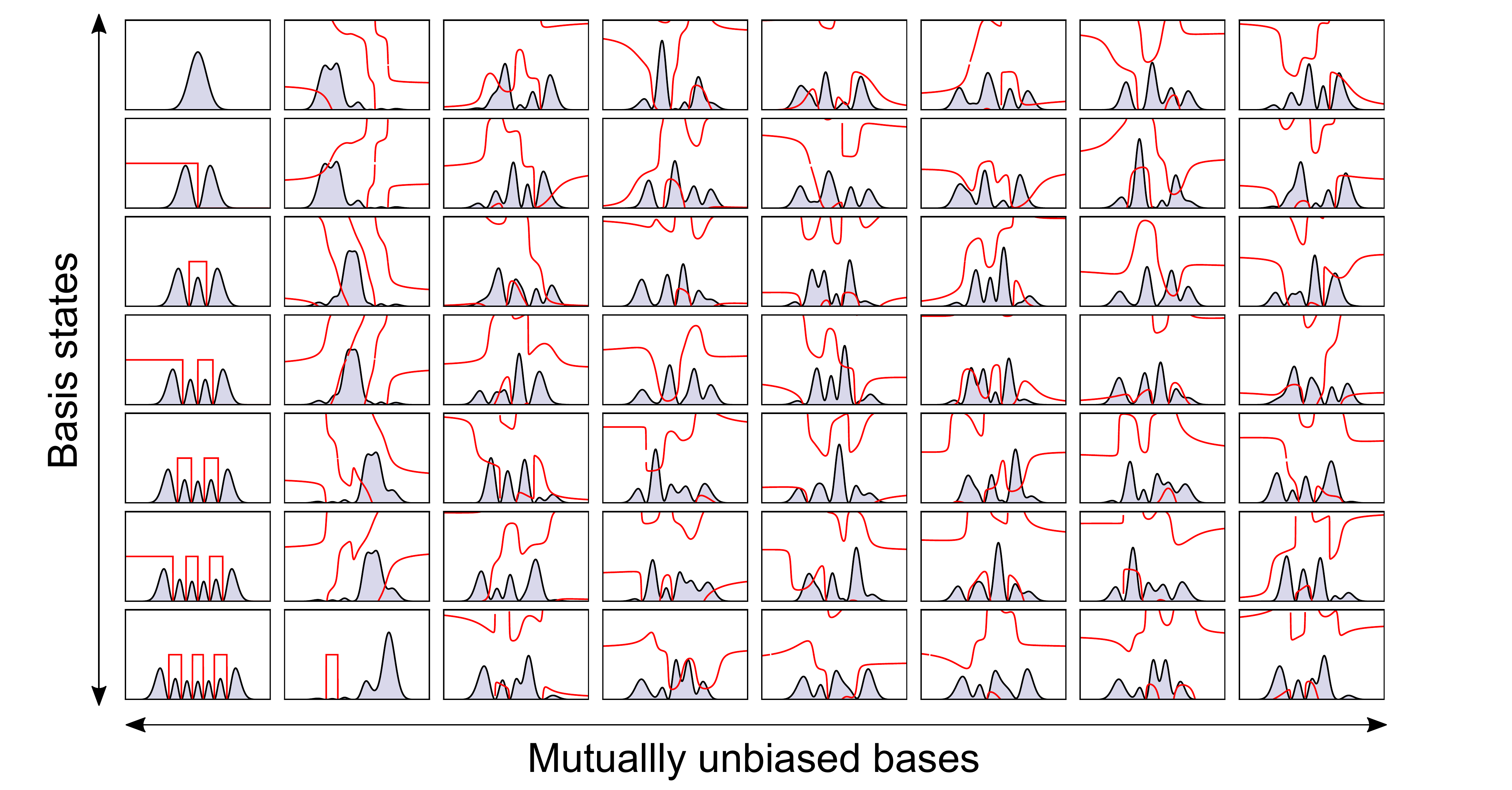}
  \end{center}
  \vspace{-0.25cm}
 \caption{\textbf{Seven-dimensional temporal-mode bases.} The spectral shapes corresponding to eight mutually unbiased seven-dimensional bases~\cite{bandyopadhyay2002new} as programmed for the reconstruction of Fig.~\ref{fig:tomoReAndIm}. The black line and blue fill correspond to the intensity $|f(\omega)|^2$, and the red line corresponds to the phase on the interval $[0,2\pi]$.}\label{fig:mub7}
\end{figure}

In this appendix, we provide technical details on the experimental setup, sketched in Fig.~\ref{fig:setup}. In Table~\ref{tab:expnumbers}, we provide measured parameters of the four PDC states explored in the main text, including the $g^{(2)}$ numbers displayed graphically in Fig.~3 of the main text. We also provide extra data detailing the purity of the PDC source as the PDC pump is chirped, as seen in Fig.~\ref{fig:chirpedsource}. In Fig.~\ref{fig:tomoReAndIm}, we provide both the real and imaginary parts of the reconstructed seven-dimensional density matrices, and compare their eigenvalues with the expected values from the JSI. In Fig.~\ref{fig:mub7}, we show the projections implemented by the QPG to reconstruct the seven-dimensional density matrices.

Our experiment is driven by an 80-MHz titanium-sapphire laser (Ti:Sa, Coherent Chameleon) and OPO (APE Compact). We create the PDC pump pulses at 769~nm by frequency doubling light from the OPO in 1~mm of bulk PPLN; the fundamental of the Ti:Sa at 876~nm is used as the QPG pump. Both pulses are shaped with approximately 0.05-nm resolution using a 4f setup consisting of a 2000~lines/mm diffracting grating, a curved mirror with a 250~mm focal length, and a reflective liquid-crystal spatial light modulator (SLM, Hamamatsu LCoS) at the focal plane~\cite{weiner2000femtosecond,monmayrant2010newcomer}. With this setup, we can directly control the bandwidth, spectral shape, and spectral phase of the pump pulses.

The PDC photons are generated through a near-degenerate type-II process in a 8-mm long PPKTP waveguide (AdvR) with a nominal poling period of 117~$\mu$m. An 80~nm broad bandpass filter is used to remove the PDC pump, and the individual photons are separated with a polarizing beamsplitter and filtered with 3~nm bandpass filters to remove side lobes. In all cases, the PDC pump energy was approximately 15~pJ per pulse, with heralded $g_h^{(2)}$s lying between $0.417\pm0.003$ for the spectrally decorrelated state and $0.246\pm0.003$ for the intensity-anticorrelated state. This relatively high production rate was used to enable reasonably precise unheralded $g^{(2)}$ measurements with 10-minute recording times. See Table~\ref{tab:expnumbers} for all $g_h^{(2)}$ values. For ease of alignment, the signal photon path can be switched for a coherent pulse from the OPO, spectrally shaped by a commercial pulse shaper (Finisar WaveShaper 4000S). The average number of generated photons can be inferred from the two-photon cross-correlation statistics~\cite{christ2011probing}, with the average generation rate of $\langle{n}\rangle\approx0.16$ for the decorrelated state deduced from a $g^{(1,1)}={\frac{1}{\langle n\rangle}+g^{(2)}}=8.303\pm0.003$.

The signal photons and the QPG pump (with an average energy-per-pulse of 250~pJ) are combined on a dichroic mirror and coupled into a 17-mm long PPLN waveguide with a poling period of 4.4~$\mu$m, fabricated in-house and designed for spatially single-mode propagation at 1540~nm. The waveguide mode of the QPG pump is imaged on a camera after the waveguide and optimized to the fundamental spatial mode. Higher-order modes produce sum-frequency signals for different time delays with central frequencies, and are filtered out of the final signal along with the second harmonic of the QPG pump by a 4f-filter. The upconverted light at 558~nm is measured on a spectrometer (Andor Shamrock SR500 spectrograph and Newton 970-BVF EMCCD camera with a 2398~lines/mm grating) to have a bandwidth of 61~pm FWHM. The 4f-filter is also used to remove spectral side lobes, which account for less than 5\% of the total upconverted photons. The upconverted green photons were detected with silicon avalanche photodiodes (SiAPDs, Excelitas), while the idler and leftover signal photons were detected with superconducting nanowire single-photon detectors (SNSPDs, PhotonSpot). All three photon paths are split into two detectors to measure photon number correlations via Hanbury-Brown-Twiss interferometry~\cite{christ2011probing}.

The joint spectral intensities (JSIs) were measured with fiber-based time-of-flight spectrometers~\cite{avenhaus2009fiber}, mapping a spectral range of 1~nm at 1540~nm to a time delay of 0.42~ns.  Assuming a flat spectral phase, the singular-value decomposition of the JSI predicts a spectral purity of 0.995 for the decorrelated JSI of Fig.~\ref{fig:tomoReAndIm}a, and 0.652 for the intensity anticorrelated JSI of Fig.~\ref{fig:tomoReAndIm}b. The marginal bandwidths (intensity FWHM) of the signal and idler photon in the decorrelated case were measured to be $4.9$~nm and $3.6$~nm, respectively.

To compensate for dispersive elements throughout the apparatus, the spectral phase of the PDC pump was optimized with the SLM to maximize the $g^{(2)}$ of the decorrelated state (Case `a'), as seen in Fig.~\ref{fig:chirpedsource}. The chirp of the phase-correlated PDC state of Fig.~\ref{fig:tomoReAndIm}c is $A=0.38\times10^6\,\mathrm{fs}^2$, where the chirp is represented as a phase in angular frequency as $\exp\left[iA(\omega-\omega_0)^2\right]$. Given a separable Gaussian PDC state with signal and idler bandwidths $\sigma_s$ and $\sigma_i$ (intensity standard deviation in $\omega$), the expected purity as a function of pump chirp $A$ is \begin{equation}P=\frac{1}{\sqrt{1+16A^2\sigma_s^2\sigma_i^2}},\label{eq:chirppurity}\end{equation} which is seen in Fig.~\ref{fig:chirpedsource} to match the experimental result well for large chirp values. While this result clearly shows that dispersion management of the pump is key for producing single-mode photons, it also provides an alternative avenue for generating highly entangled photon pair states. For tasks requiring highly multimode photons, this method of increasing the number of modes present can make use of the entire PDC pump bandwidth, and therefore does not significantly affect the pair generation rate of the source in power-limited situations.

The central wavelength and time delay of the QPG pump relative to the PDC signal photons were set by optimizing the ratio of upconversion between HG0 and HG1 projections. The spectral phase and bandwidth of the QPG pump were adjusted to maximize the visibility between HG0 and HG2 projections. Pulse bandwidths as measured on a spectrometer (Andor Shamrock SR500 with a 1200~l/mm grating) are given in Table~\ref{tab:expnumbers}.

For the $g^{(2)}$ measurements of Fig.~4 of the main text, the QPG is effectively set to `OFF' by delaying the pump by 5~ps, where it does not interact with the PDC photons. The QPG pump is delayed rather than blocked in order to ensure all three measurements are subject to the same background noise, which may arise from scattering of the transmitted QPG pump or a broadband parametric noise from errors in periodic poling~\cite{pelc2011long}. With all laser pulses blocked, the ambient and detector dark-count rate was approximately 1.8k-per-detector-per-second for the SNSPDs and 350-per-detector-per-second for the APDs. Coincidences are registered within a 3~ns window, and the expected dark counts outside this window are subtracted. When the PDC pump is blocked but the QPG pump is coupled through the system, we measure extra background counts of approximately 4.9k and 80 counts-per-detector-per-second on the SNSPDs and APDs, respectively. This has a negligible effect on the measurements of the upconverted photons, but significantly impacts the $g^{(2)}$ of the transmitted PDC signal photons, as seen by comparing the ``QPG pump blocked'' and the ``QPG pump delayed'' $g^{(2)}$ values in Table~\ref{tab:expnumbers}. Note that no background subtraction is employed for the tomography results, but they are conditioned upon coincidence with an idler detection. When measuring in coincidence, the background rate measured by the APDs drops below 8-per-detector-per-second, while the coincidence detection rate for the Gaussian projection onto the single-mode state is over 7000-per-detector-per-second, providing a more-than-satisfactory signal-to-noise.


\begin{table}
\begin{tabular}{|l|c|c|c|c|}
\hline
Reference & (a) & (b) & (c) & (d) \\ \hline
\hline
PDC Pump Shape & HG0 &  HG0 &  HG0 &  HG1 \\ \hline
PDC Pump Bandwidth & 1.72~nm &  0.54~nm &  1.49~nm &  1.31~nm \\ \hline
PDC Pump Chirp & 0 &  0 &  $0.38\times10^6\,\mathrm{fs}^2$ &  0 \\ \hline
QPG Pump Bandwidth & 1.54~nm &  1.05~nm &  1.58~nm &  1.30~nm \\
\hline\hline
Purity of $\rho$ from reconstruction & $0.896\pm0.006$ & $0.523\pm0.008$ & $0.317\pm0.005$ & $0.531\pm0.004$ \\
\hline
Expected purity from JSI & $0.995$ & $0.652$ & $0.377$* & $0.542$* \\
\hline\hline
Transmitted~$g^{(2)}$, QPG pump blocked & $1.929\pm0.008$ & $1.528\pm0.010$ & $1.327\pm0.005$ & $1.498\pm0.006$ \\
\hline
Transmitted~$g^{(2)}$, QPG pump delayed & $1.861\pm0.003$ & $1.494\pm0.003$ & $1.302\pm0.002$ & $1.461\pm0.003$ \\
\hline
Transmitted~$g^{(2)}$, QPG pump HG0 & $1.827\pm0.004$ & $1.456\pm0.004$ & $1.277\pm0.002$ & $1.467\pm0.003$ \\
\hline
Transmitted~$g^{(2)}$, QPG pump HG1 & $1.875\pm0.003$ & $1.512\pm0.004$ & $1.308\pm0.002$ & $1.446\pm0.003$ \\
\hline\hline
Upconverted~$g^{(2)}$, QPG pump HG0 & $1.975\pm0.015$ & $2.044\pm0.037$ & $1.983\pm0.026$ & $1.949\pm0.033$ \\
\hline
Upconverted~$g^{(2)}$, QPG pump HG1 & $2.078\pm0.194$ & $1.951\pm0.105$ & $1.925\pm0.063$ & $1.993\pm0.025$ \\
\hline
\hline
Transmitted~$g_h^{(2)}$, QPG pump delayed & $0.417\pm0.003$ & $0.246\pm0.003$ & $0.374\pm0.002$ & $0.393\pm0.003$ \\
\hline
Upconverted~$g_h^{(2)}$, QPG pump HG0 & $0.423\pm0.005$ & $0.319\pm0.009$ & $0.501\pm0.011$ & $0.572\pm0.017$ \\
\hline
\end{tabular}
\caption{Pump bandwidths and measured $g^{(2)}$s for the four PDC states explored in the main text, corresponding to the JSIs of Fig.~\ref{fig:tomoReAndIm}. The error of the purity from the tomographically reconstructed density matrices $\rho$ are found through Monte Carlo simulation assuming the coincidences measured have Poissonian error. The expected purity from the JSIs correspond to the singular value decomposition assuming a flat phase, except in cases marked (*) where faithful implementation of the intended phase is assumed. All $g^{(2)}$ values are corrected for detector dark counts assuming a 3~ns coincidence window. $g^{(2)}_h$ is the heralded second-order correlation function, which is zero for the ideal single-photon Fock state and one or greater for all classical states of light.}
\label{tab:expnumbers}
\end{table}

\end{document}